\documentclass[aip,reprint]{revtex4-1}
\usepackage[CJKbookmarks]{hyperref}
\usepackage{amsmath,amssymb,esint} 
\usepackage{mathtools}
\allowdisplaybreaks[1]
\usepackage{graphicx}
\usepackage{tikz}
\usetikzlibrary{external}
\renewcommand*{\vec}[1]{\mathbf{#1}} 
\newcommand{\dif}{\,\mathrm d}
\newcommand\mi{\mathrm{i}}
\newcommand\e{\mathrm{e}}

\DeclareMathOperator\imag{Im}
\DeclareMathOperator\real{Re}
\DeclareMathOperator\diag{Diag}

\newcommand{\modi}[1]{#1}
\graphicspath{{figs/}}
\begin{document}
\title{Correction to the Effective Refractive Index and the Confinement Factor 
in Waveguide Modeling for Quantum Cascade Lasers}
\author{Ming Lyu}
\email{minglyu@princeton.edu}
\author{Claire Gmachl}
\affiliation{Department of Electrical Computer Engineering, Princeton University, 
Princeton, NJ, 08544, USA}
\date{\today}
\begin{abstract}
The equations for the effective medium refractive index and for the 
confinement factor in the waveguide design for 
quantum cascade lasers are derived.
Compared to equations used in prior literature, 
by applying rigorous perturbation theory and
including the effect of the anisotropic optical gain and non-Hermitian 
properties of the waveguide structure and materials,
a few percent correction should be made to the confinement factor and 
the effective gain. 
This result can easily be generalized to any optical devices with 
a layered structure. 
\end{abstract}
\maketitle
\section{Introduction}
Active semiconductor optical devices including LEDs, lasers, meta-material 
devices, etc. have been developing rapidly, introducing complex multi-layered 
optical structures.
Concepts including the effective refractive index and the confinement factor
\cite{anderson1965mode,hakki1975gain,botez1978analytical} are often used
to simplify the modeling of those structures. One good example is 
quantum cascade lasers (QCLs) \cite{faist1994quantum}, where tens-of-atomic scale 
layers as quantum wells and multi-layered sub-wavelength optical claddings are 
built on 
a single wafer to produce efficient lasers of mid-infrared to THz light. 

Since the invention of QCLs, much effort has been made to improve 
the laser performance, both via active region design and the waveguide
design. Different waveguiding mechanisms including index guiding, 
plasmonic guiding, and double-metal waveguiding \cite{firstplasmon,doublemetal}
are widely used to reduce the optical loss of the 
device as well as to increase the confinement factor. 

The confinement factor in particular has been defined differently in different 
references \cite{anderson1965mode,hakki1975gain,botez1978analytical,modeling,
adams1981introduction,huang1996}%
. 
To the best of our knowledge, there is not any published 
analytical analysis about what should be the more accurate equation for
the effective medium refractive index and the confinement factor that takes 
into consideration the polarization selection law for QC gain, 
the particulars of QCL layer structures and the 
non-Hermitian property of lossy materials in the waveguide. 

In this work, we derive the equations for the effective 
refractive index and the confinement factor directly from Maxwell's equations, 
and discuss when conventional expressions often used in literature may lead to 
noticeable errors. 

In Section~\ref{sec:def} we define the model and the variables for the work;
running wave effective refractive index in an infinite periodical structure
is investigated in Section~\ref{sec:running} to derive the effective medium 
refractive index in active core of a QCL, which is compared with numerical 
result to show the validity of the effective medium approximation 
on typical parameters; 
In Section~\ref{sec:analytic} and \ref{sec:perturb} we show the analytical 
and perturbative treatment of the guided mode in a 2D waveguide respectively, 
where the linear perturbation gives the confinement factor. 

\section{1D Maxwell's Equations for a 2D Waveguide}\label{sec:def}
\begin{figure}[!htp]
\begin{minipage}[c]{0.35\linewidth}
\vspace{0.5em}


\includegraphics{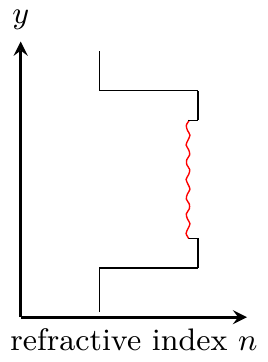}
\end{minipage}%
\begin{minipage}[c]{0.65\linewidth}
\centering




\includegraphics{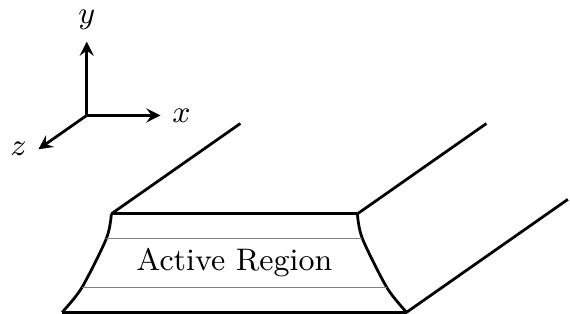}
\end{minipage}
\caption{The coordinate system: for sketch purposes we draw a ridge 
waveguide to show the direction of wave propagation. The 2D waveguide is a good 
approximation when the ridge width along $x$ is much larger than the wavelength.
The refractive index profile varies in different waveguide designs, but usually 
includes cladding layers and periodical active layers (red). 
\label{fig:coordinate}}
\end{figure}
In the following context we assume relative permeability $\mu = 1$;
the relative permittivity $\varepsilon = n^2$ is a function of $y$, which is 
the growth direction of the epi-layers;
the structure is constant and infinite in 
$x$ and $z$ direction, where the $z$ direction is the direction 
of wave propagation; see Fig.~\ref{fig:coordinate} for a diagram of the 
coordinates. 

Maxwell's equations at a frequency $\omega$ can be written as:
\begin{align}
    & \nabla \times \frac1{\varepsilon} \nabla \times \vec H 
     = \frac{\omega^2}{c^2} \vec H \\
    & \vec E 
    = \frac{\mi}{\omega\varepsilon_0 \varepsilon} \nabla\times \vec H
\end{align}
where $\varepsilon$ generally should be a symmetric tensor with complex elements, 
but we assume it to have a principle axis along the $x$, $y$ and $z$ direction,
noted as $\varepsilon = \diag\{\varepsilon_x, \varepsilon_y, \varepsilon_z\}
=\diag\{n_x^2, n_y^2, n_z^2\}$. This is justified by the $y$-axial symmetry 
of the structure, and the fact that growth and fabrication most commonly happens
along major crystal directions. 

Here, since $\partial_z = \mi\beta k$ ($\beta$ is the effective 
refractive index of the guided mode), $\partial_x = 0$, the Maxwell's
equation gives
\begin{widetext}
\begin{align}
    &\nabla\times\frac 1\varepsilon\nabla\times \vec H = 
    \begin{pmatrix}
        -\partial_y \varepsilon_z^{-1} \partial_y + \beta^2 k^2\varepsilon_y^{-1}
        & 0 & 0\\
        0 & \beta^2 k^2\varepsilon^{-1}_x & 
        \mi \beta k\varepsilon^{-1}_x\partial_y \\
        0 & -\mi \beta k\partial_y \varepsilon^{-1}_x
        & \partial_y\varepsilon^{-1}_x\partial_y
    \end{pmatrix}\begin{pmatrix}
        H_x \\ H_y \\ H_z
    \end{pmatrix}
\end{align}
\end{widetext}
which is naturally block diagonal, giving modes $H_y=H_z=0$ 
(transverse magnetic, TM) and $H_x=0$ (transverse electric, TE). 
For the TM modes, which is the mode of QCLs due to the selection law 
for intersubband transitions \cite{faist1994quantum}, 
the 3D equation reduces to 1D:
\begin{align}
    &\left(-\frac{\partial}{\partial y} 
    \frac 1{n^2_z}
    \frac{\partial}{\partial y} 
    + \frac{\beta^2 k^2}{n^2_y}\right)
    H_x = \frac{\omega^2}{c^2} H_x \label{eq:hx}\\
    & E_y = -\frac{\beta k H_x}{\omega\varepsilon_0n_y^2} 
    \qquad E_z = -\frac{\mi}
    {\omega\varepsilon_0n^2_z}\frac{\partial H_x}{\partial y}
\end{align}

\section{The Unguided Effective Refractive Index in The Active Region}
\label{sec:running}
The active region of QCLs typically consists of tens of periods 
of active and injection layers, consisting of multiple quantum wells and 
barriers, each of which are typically a few atoms thick, 
adding up to a period length of a few hundred angstroms. 
This period is about one order of magnitude smaller than the wavelength in 
vacuum, and therefore the effective medium theory is commonly applied.
In this section we show, however, a more accurate expression for the effective
refractive index for QCL active regions. 

For an active region with period $L_p$, 
assuming a structure with infinite number of periods, 
the Bloch theory gives $H_x = u(y)\e^{\mi k y}$, 
with $-\pi/L_p < k \le \pi/L_p$ and $u(y+L_p) = u(y)$. In the frequency
domain Eq.~(\ref{eq:hx}) is
(for simplicity here we use the fact that within each individual 
quantum well and quantum barrier layer
the material is isotropic, i.e. $n_x = n_y = n_z = n$): 
\begin{align}
    &u(y) 
    = \sum_{j=-\infty}^{\infty} u_j \e^{\mi 2\pi j y / L_p} \\
    &\frac{1}{n(y)^2}
    = \sum_{j=-\infty}^{\infty} \frac 1{n_j^2} \e^{\mi 2\pi j y / L_p}\\
    &\sum_q \left[
    \left(\frac{2\pi j}{L_p} + k\right) 
    \left(\frac{2\pi q}{L_p} + k\right)
    + \beta^2 k^2\right] \frac{u_q}{n_{j-q}^2}
    = \frac{\omega^2}{c^2}u_j \label{eq:frequ}
\end{align}
where $u(y)$ is the slowly varying amplitude of the field, $n(y)$ is the 
spatial dependent refractive index, $u_j$ and $1/n_j^2$ are Fourier series
of $u(y)$ and $1/n(y)^2$, $\beta$ is the effective refractive index of the
waveguide as defined in Eq.~(\ref{eq:hx}), $k = 2\pi/\lambda$ is the wave vector 
amplitude in vacuum. \modi{Eq.~(\ref{eq:frequ}) is the equation of the Fourier 
components of Eq.~(\ref{eq:hx})}

When $kL_p \ll 1$, $u(y)$ varies slowly at the $L_p$ scale and $u_q \approx 0$ 
for $|q|\neq 0$. The effective medium result comes with the approximation that
$u(y)\approx u_0$, which leads to the effective refractive as the zero frequency 
component of the refractive index profile:
$n_{\text{TM}} = \langle 1/n^2\rangle^{-1/2}$ or $\varepsilon^{-1}_{\text{TM}} =
\langle \varepsilon^{-1} \rangle$, where $\langle\bullet\rangle$ means average 
value weighted by the layer thickness. 
Similarly for the TE mode the result is 
$n_{\text{TE}} = \langle n^2\rangle^{1/2}$ 
or $\varepsilon_{\text{TE}} = \langle \varepsilon \rangle$.

\begin{figure*}[!htp]
\begin{minipage}[t]{0.5\linewidth}
\centering
\includegraphics[width=\linewidth]{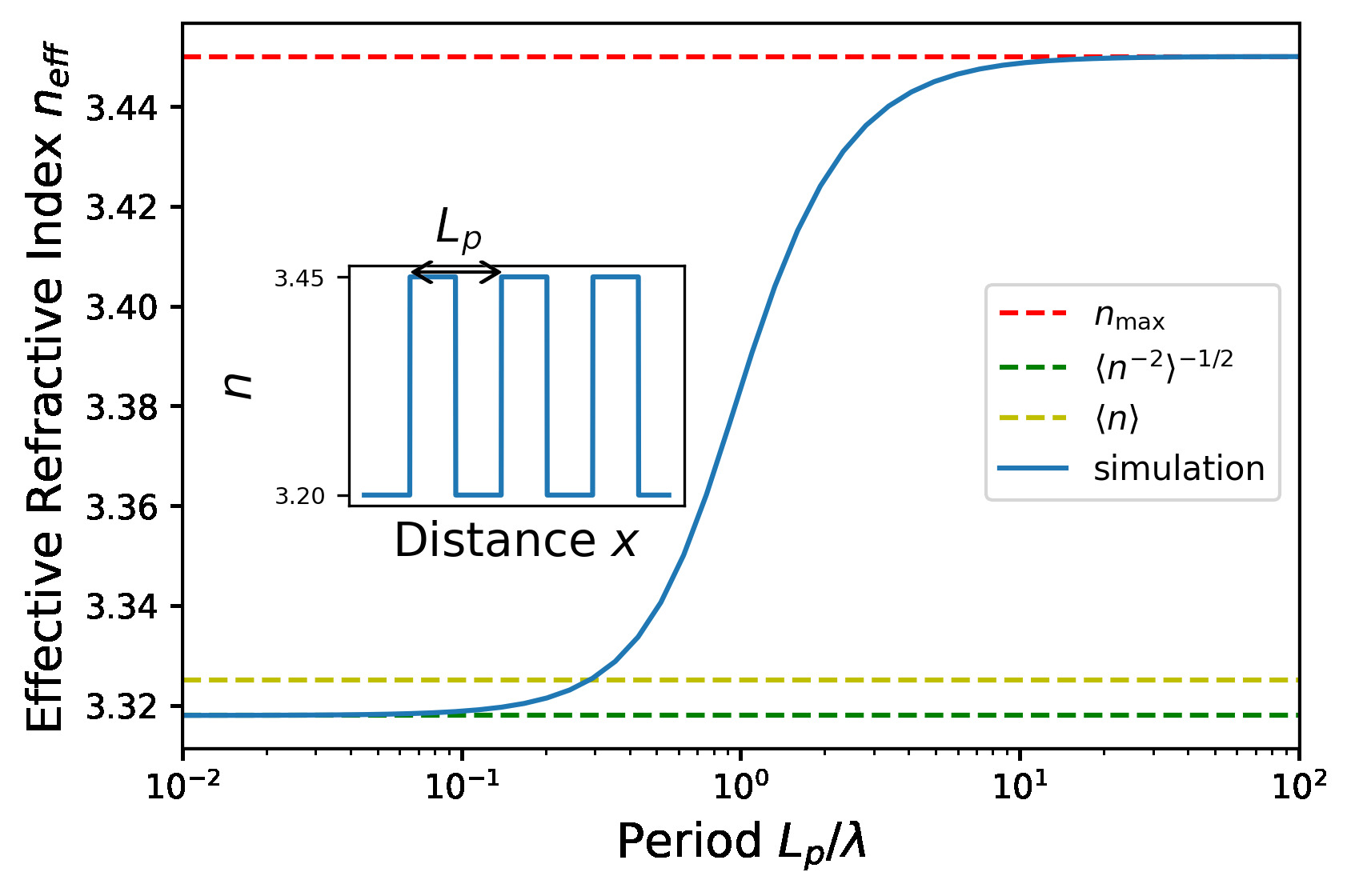}
\end{minipage}%
\begin{minipage}[t]{0.5\linewidth}
\centering
\includegraphics[width=\linewidth]{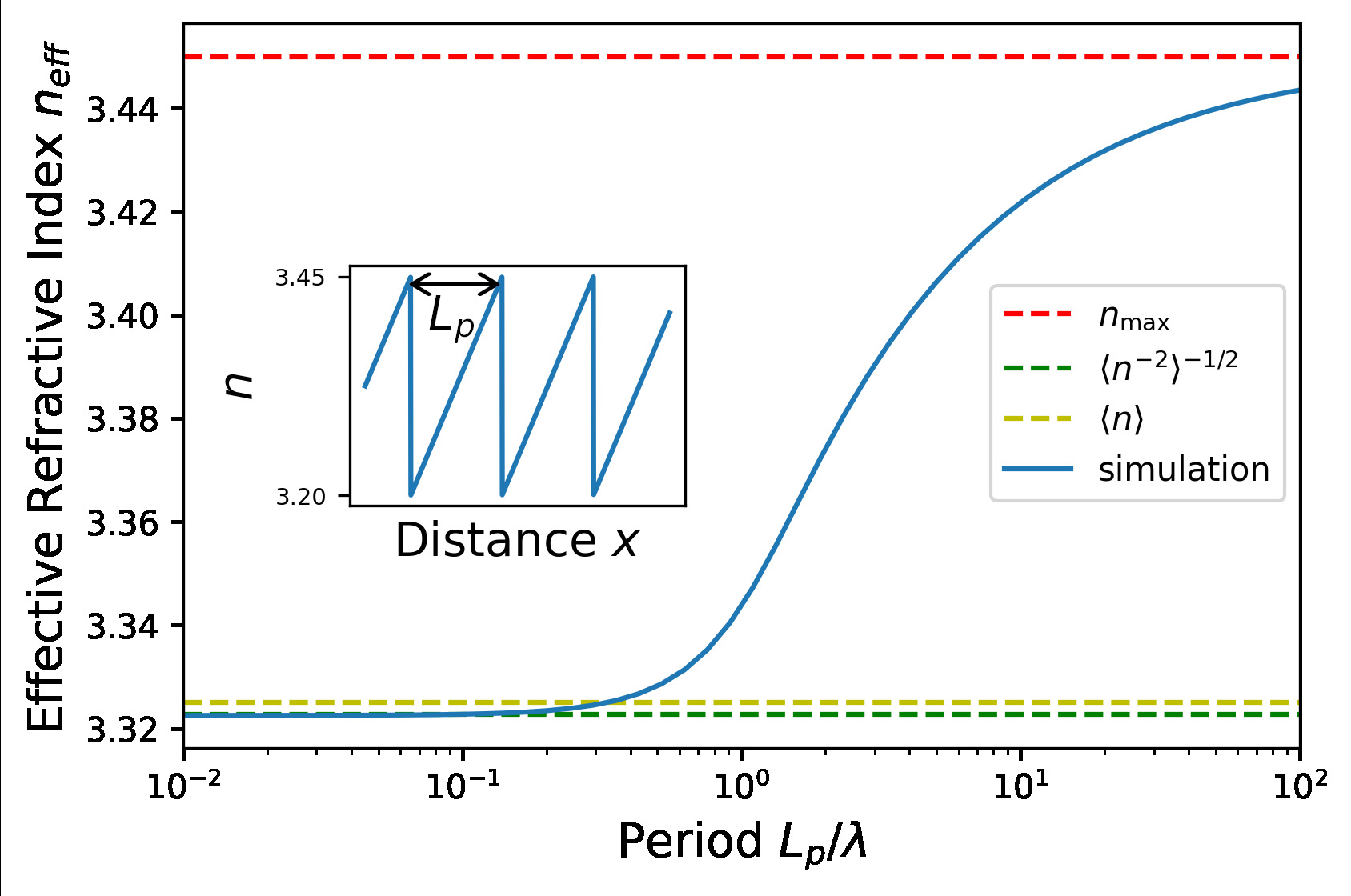}
\end{minipage}
\caption{Simulation of the effective refractive index of different refractive 
index profiles (insets, \textbf{left} alternating two different 
materials with the same thickness; 
\textbf{right} a ratchet shaped refractive index as a simplified model 
for increasing the relative ratio of high refractive index
material in a QC structure).
$L_p$ is the period of the profile, $\lambda$ is the wavelength in vacuum. 
\label{fig:periodicref}}
\end{figure*}

This result is very similar with the well-known effective medium result for 
different polarizations $\varepsilon_\parallel = \langle \varepsilon \rangle$ 
and $\varepsilon_\perp = \langle \varepsilon^{-1}\rangle^{-1}$ \cite{cai2010optical}
of a birefringent material,
except that it is for the TE and TM modes, rather than for the electric field 
of different directions. It is worth noting that for the TM mode there are 
non-zero electrical field components in both parallel ($z$) and perpendicular 
($y$) directions. This difference becomes noticeable when in the following 
we consider in more detail the anisotropic refractive index in a 1D waveguide, 
induced by the near-atomic-level layering of different semiconductor materials
in the active region (material refractive index examples are shown in the 
insets in Fig.~\ref{fig:periodicref}).

In Fig.~\ref{fig:periodicref} we compare the result of exact solution of 
Eq.~(\ref{eq:frequ}) (by diagonalizing the linear operator on $u_q$ in the 
left-hand-side up to a large enough cutoff) for the fundamental mode 
and the result of effective medium
theory for two different periodic refractive index profiles, 
where we can see that: (a) for $L_p\lesssim 0.1\lambda$ (where $\lambda$ is 
the wavelength in vacuum) the effective medium theory is a very
good approximation; (b) in the small wavelength limit ($L_p/\lambda \to \infty$), 
the result reduces to simple index guiding in the large refractive index 
region, so $n_{\text{eff}} = n_{\text{max}}$; (c) if the 
effective refractive index were calculated from arithmetic averaging 
$n_\text{eff} =\langle n \rangle$, it would lead to $\lesssim 0.5\%$ error;
considering relatively small refractive index contrast in many photonic 
structures, this can be non-negligible.

\section{Guided Mode Calculation with the Transfer Matrix Method}
\label{sec:analytic}
In QCLs as well as in conventional diode lasers, the waveguide claddings 
are typically implemented with several layers of
different refractive index materials of sub-wavelenth thickness. 
Such structure can be analytically solved using the transfer matrix method, 
as in \cite{Chilwell:84}. 

Here we adopt the method for anisotropic materials (this can either be the 
layering-induced anisotropy discussed in the previous sections or material 
anisotropy), for the purpose of discussing the anisotropic gain/loss in QCLs.
This is necessary because in the active region of a QCL, the gain is only on 
the electrical field in 
$y$ direction due to confined dipole direction, and the plasmonic loss is only 
in the $x$-$z$ plane due to discrete quantum levels in $y$.

Eq.(\ref{eq:hx}) can then be written as: 
\begin{equation}
    n^2_y\frac{\partial}{\partial y} \frac{1}{n^2_z}\frac{\partial}{\partial y}
     H_x =
    -(n^2_y -\beta^2 )k^2 H_x
    \label{eq:linear_hx}
\end{equation}
The equation naturally suggests interface conditions by requiring 
$H_x$ and $(1/n^2_z)\partial H_x/\partial y$ to be continuous. 
This is consistent with the electrical field interface condition, which requires 
that $D_y = \varepsilon_y\varepsilon_0 E_y = - H_x \beta k/\omega$ and 
$E_z = [(1/n^2_z)\partial H_x/\partial y] / (\mi\omega\varepsilon_0)$ are continuous. 

Within the same layer where $n_{y, z}$ are constant, 
\begin{align}
&H_x (y) = H_x^+\e^{\mi\alpha y} + H_x^-\e^{-\mi\alpha y} \\
&E_z (y) = \gamma(H_x^+\e^{\mi\alpha y} - H_x^-\e^{-\mi\alpha y}) 
\label{eq:ezhxpm}\\
&\gamma\equiv \frac{\alpha}{kn^2_z} \sqrt{\frac{\mu_0}{\epsilon_0}}
\label{eq:gamma}\\
&\alpha \equiv k\sqrt{n_z^2-\frac{\beta^2}{n_y^2}}\label{eq:alpha}
\end{align}
where $H_x^+$ and $H_x^-$ are the positive and negative $y$-propagating 
component of the magnetic field in $x$ direction $H_x$, $\alpha$ is the 
$y$ component of the effective wave-vector, 
$\gamma$ is the effective wave impedance. 

The transfer matrix $M_L$ for a layer with thickness $L$ is given by:
\begin{widetext}
\begin{equation}
    \begin{pmatrix}
        E_z(0) \\
        H_x(0)
    \end{pmatrix} = 
    \begin{pmatrix}
        \cos\alpha L & -\mi\gamma\sin\alpha L \\
        -\mi\gamma^{-1}\sin\alpha L & \cos\alpha L
    \end{pmatrix}
    \begin{pmatrix}
        E_z (L) \\
        H_x (L)
    \end{pmatrix} \equiv 
    M_L
    \begin{pmatrix}
        E_z (L) \\
        H_x (L)
    \end{pmatrix} 
    \label{eq:single_tranfer}
\end{equation}
\end{widetext}
For complex valued $\beta$ and $n_{y,z}$, 
the square root in Eq.~(\ref{eq:alpha}) is double-valued, 
but this does not affect the matrix $M_L$ because all elements
in the matrix are even functions of $\alpha$. 
However, this double-value will affect the boundary condition for a guided 
mode, as we will show in the following. 

Let the transfer matrix for $i$-th layer be $M_i$. 
The transfer matrix for the whole structure is a matrix product of all $M_i$-s: 
$M = \prod M_i = M_1M_2\cdots M_N$. 
For a guided mode the field decays before the first and after the last layer,
which gives the boundary condition
$E_z(0^-) = -\gamma_0H_x(0^-)$ and $E_z(L^+) = \gamma_sH_x(L^+)$ by 
choosing only $H_x^+$ or $H_x^-$ in Eq.~(\ref{eq:ezhxpm}). This means 
$(\gamma_0, 1)^T$ is parallel to $M(\gamma_s, 1)^T$, or:
\begin{equation}
    \chi_M(\beta) \equiv 
    \gamma_sM_{11} + M_{12} + \gamma_s\gamma_0M_{21} + \gamma_0 M_{22} = 0
\end{equation}
where $\gamma_s$ and $\gamma_0$ are, respective, the $\gamma$-s in 
Eq.~(\ref{eq:gamma}) for the substrate after the last layer and 
for the environment before the first layer, and choosing the branch of the
square root to have positive imaginary part $\imag\alpha > 0$; 
$M_{ij}$ is the $i$-th row, $j$-th column element of the matrix $M$. 
$\chi_M$ is called the modal-dispersion function.
The modal-dispersion function transforms the eigen-problem Eq.~(\ref{eq:hx}) 
in function space to a root-finding problem. 

The formula is applicable for both index guiding and for plasmonic guiding 
because the refractive index in the equations can be complex. 
For plasmonic guiding the only difference is that there should be a layer with 
refractive index with large imaginary part. 

The above equations give an algorithm to calculate the effective refractive 
index for guided modes in any layered 2D waveguide, including the anisotropic 
effect (difference of $n_y$ and $n_z$ in Eq.~(\ref{eq:linear_hx}) -- (\ref{eq:alpha}))
and non-unitary material (loss/gain with complex refractive index). 

\section{Perturbation Theory and Confinement Factor}
\label{sec:perturb}
In principle, we can calculate the effective gain $g_{\text{eff}}$ 
of a guided mode directly from the last section, 
as it is proportional to the negative imaginary
part of the waveguide effective refractive index. 
However, to simplify the modeling for threshold current and slope efficiency,
a linear response form for the gain is preferred, which is discussed in this
section, which also introduces the concept of ``confinement factor''. 
The linear response is the result of perturbation theory \cite{coldren2012diode},
and the confinement factor appears as the ratio of 
the waveguide gain and the material gain. 
However, there is not much previous work of a rigorous mathematical derivation 
for the formula for the confinement factor \cite{huang1996}, and some
misunderstanding has been ignored. 

\begin{figure*}[!htp]
\begin{minipage}[t]{0.5\linewidth}
\centering
\includegraphics[width=\linewidth]{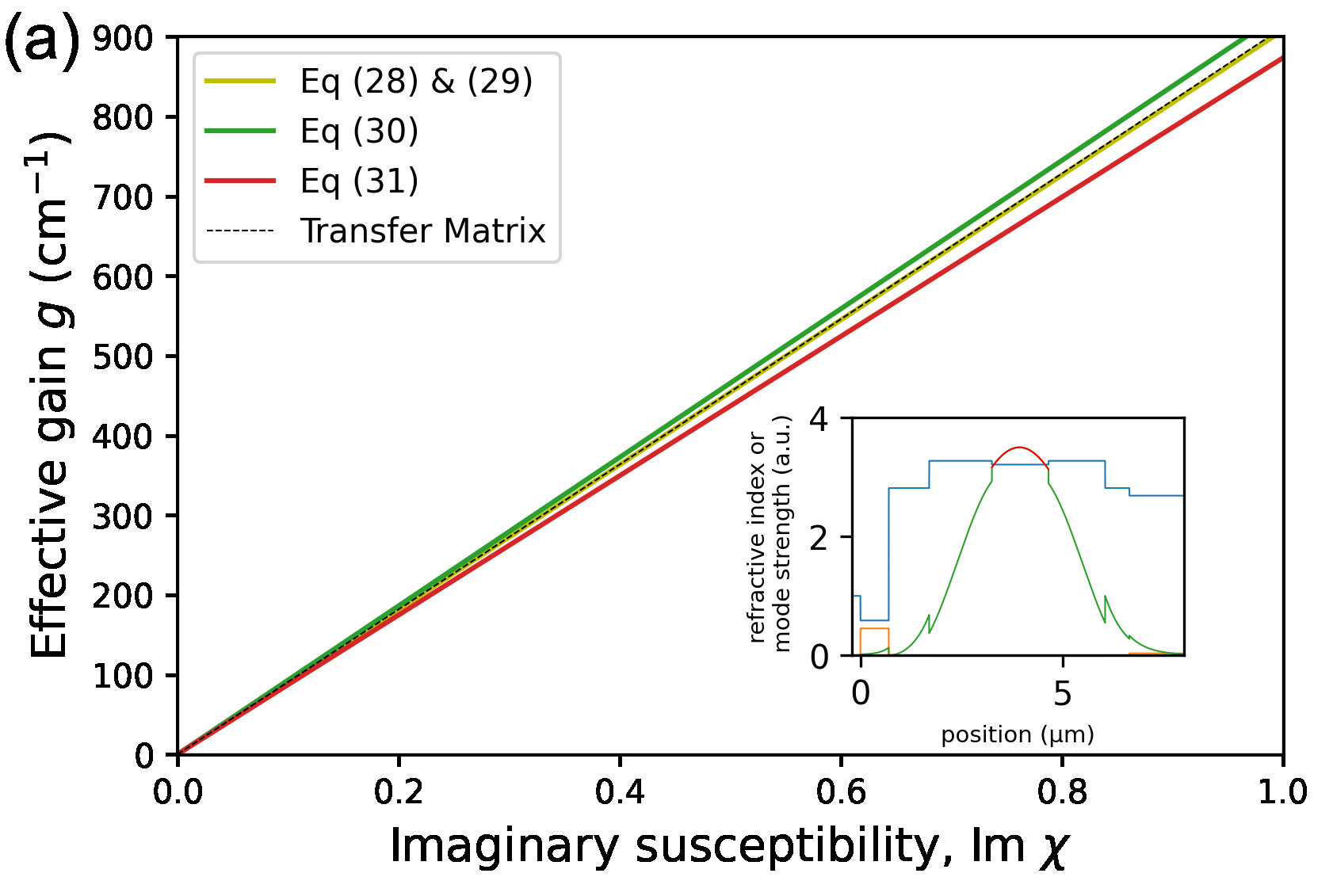}
\end{minipage}%
\begin{minipage}[t]{0.5\linewidth}
\centering
\includegraphics[width=\linewidth]{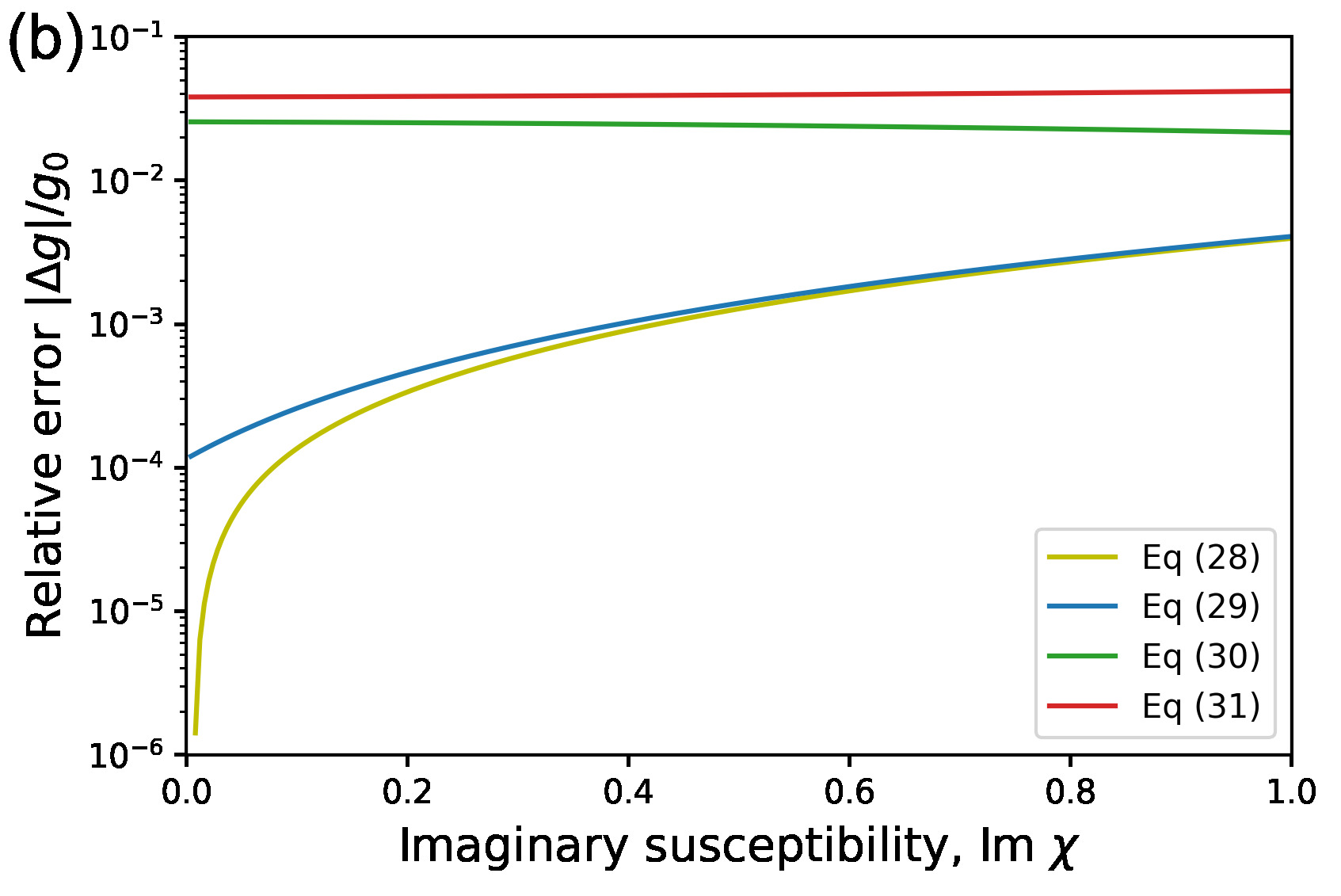}
\end{minipage}\\
\begin{minipage}[t]{0.5\linewidth}
\centering
\includegraphics[width=\linewidth]{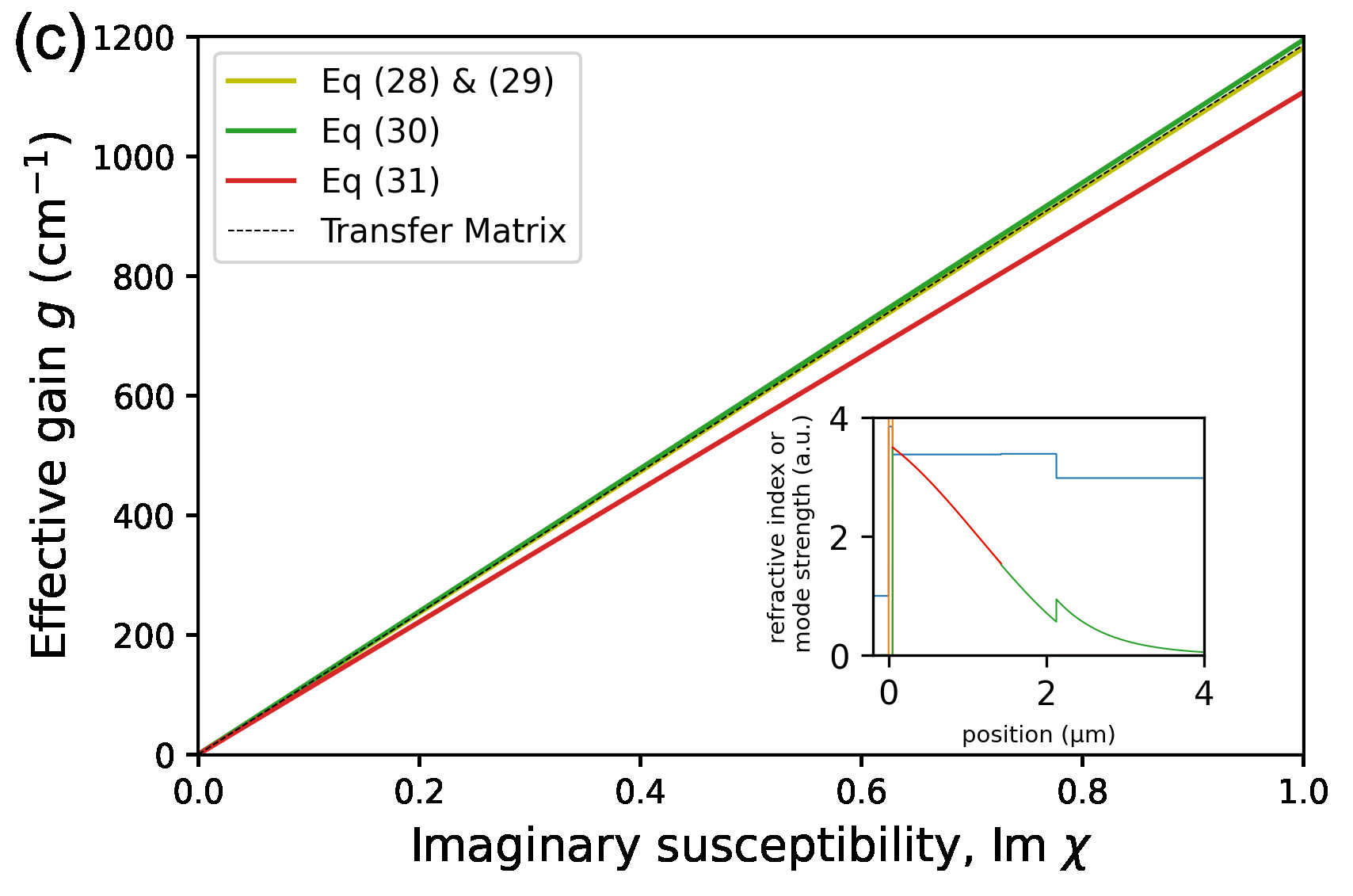}
\end{minipage}%
\begin{minipage}[t]{0.5\linewidth}
\centering
\includegraphics[width=\linewidth]{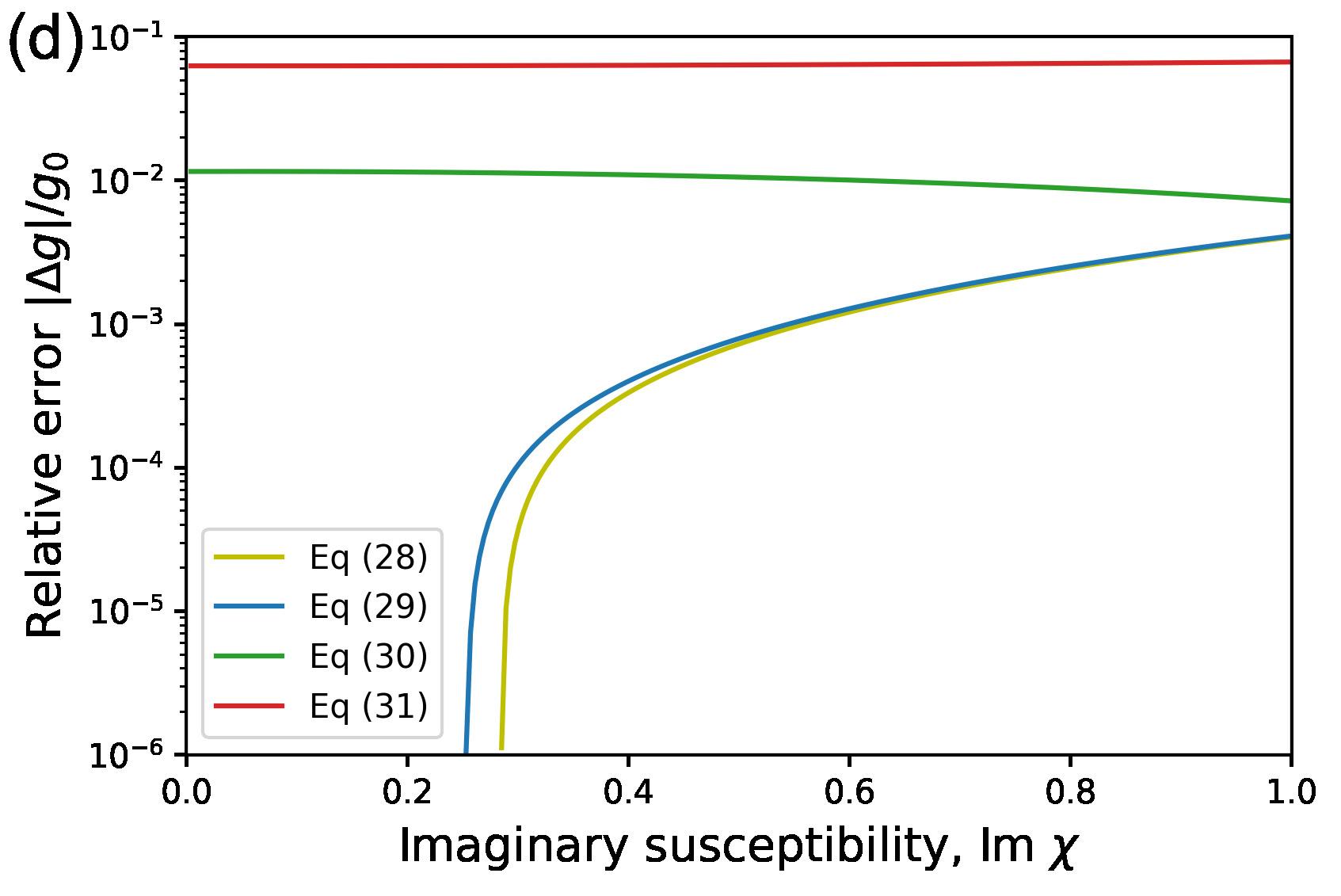}
\end{minipage}
\caption{The effective optical gain of the waveguide vs the imaginary 
susceptibility of the QC active material (\textbf{(a)} and \textbf{(c)}) 
and the relative error of the linear estimation (\textbf{(b)} and \textbf{(d)}),
calculated from Eq.~(\ref{eq:exactgain}) and
the confinement factor defined in 
Eq.~(\ref{eq:confinementreal})--(\ref{eq:confinement_e}). 
The non-perturbed result (dashed black line) from the transfer matrix method is 
considered the exact result.
\text{Insets} are the waveguide structure for the calculation
(\textbf{(a)} and \textbf{(b)}: \cite{firstgaas}; 
\textbf{(c)} and \textbf{(d)} \cite{firstplasmon}), 
where the blue lines and orange lines are the real and imaginary part of 
the refractive index respectively; 
green lines are the mode strength in arbitrary units, 
with the field in the active region colored red. 
\modi{In \textbf{(a)} and \textbf{(c)} the Eq.~(28) and (29) curves are grouped 
because the difference is too small so that they overlap in the figure.
A demonstration case where these two equations are noticeably different is shown 
in Fig.~\ref{fig:alternating-gain}.}
\label{fig:device-gain}}
\end{figure*}

In this section we modify the traditional perturbation method for Maxwell's 
equation for non-Hermitian materials, and derive the equation for 
the confinement factor from first principle. 
Compared to previous work \cite{huang1996}, our method is compatible with 
complex refractive indices (for amplifying and lossy material) and
the anisotropic property of the QC layers.

The standard perturbation theory for eigen-problems relies on the property of
a Hermitian operator, but for the eigen-problem in Eq.~(\ref{eq:hx})
\begin{equation}
    \beta^2 H_x = \Theta H_x \equiv n_y^2\left(
    \frac{\partial}{\partial y}\frac{1}{n^2_z}\frac{\partial}{\partial y} 
    + k^2\right)H_x
\end{equation}
the operator $\Theta$ is not Hermitian under the most commonly used 
inner product ($\langle A_1, A_2\rangle = \int A_1 A_2\dif y$)
due to the position dependence of $n_y$ and due to the imaginary part of 
$n_y$ and $n_z$.
However, if we define a pseudo-inner product (pseudo because it is not 
positive definite) as: 
\begin{equation}
    \langle A_1, A_2\rangle \equiv \int\frac{1}{n^2_y}A_1 A_2\dif y
\end{equation}
the operator $\Theta$ is ``Hermitian'' for a guided (modified from bounded) mode: 
$\langle A_1, \Theta A_2\rangle = \langle \Theta A_1, A_2\rangle$.
With such an inner product, we can build a perturbation theory on 
$\Theta + \delta\Theta$:
\begin{widetext}
\begin{align}
    &(\Theta + \delta \Theta) (H_x + \delta H_x + \mathcal O(\delta^2))
    = (\beta^2 + \delta\beta^2 + \mathcal O(\delta^2))
    (H_x + \delta H_x + \mathcal O(\delta^2)) \\
    &\Rightarrow
    \delta\beta^2 = \frac{\langle H_x, 
    \delta\Theta H_x\rangle}{\langle H_x, H_x\rangle}
\end{align}
when $\delta\Theta$ corresponds to a change in refractive index $\delta n$, 
\begin{align}
    \langle H_x, H_x\rangle &= \int \frac{1}{n_y^2} H_x^2 \dif y
    = \frac{\omega^2\varepsilon_0^2}{\beta^2}\int n^2_y E_y^2\dif y\\
    \langle H_x, \delta\Theta H_x\rangle &= 
    \int\, H_x \left[\frac{\delta n^2_y}{n^2_y}\frac{\partial}{\partial y}
    \frac{1}{n^2_z}\frac{\partial}{\partial y} + 
    \frac{\partial}{\partial y}\delta\frac{1}{n^2_z}\frac{\partial}{\partial y} + 
    \frac{\delta n^2_y}{n_y^2}k^2\right] H_x \dif y\\
    &= \int \frac{\delta n_y^2}{n^4_y}H_x \Theta H_x\dif y + \int H_y
    \frac{\partial}{\partial y}\delta\frac{1}{n^2_z}\frac{\partial}{\partial y}
    H_x\dif y\\
    &= \int\delta n_y^2 \left(\frac{\beta}{n^2_y} H_x\right)^2 \dif y - 
    \int\delta\frac{1}{n^2_z}\left(\frac{\partial}{\partial y}H_x\right)^2\dif y\\
    &= \omega^2\epsilon_0^2\left(\int\delta n^2_yE_y^2\dif y +
    \int n_z^4 \delta\frac{1}{n^2_z} E_z^2\dif y\right)
    \approx \omega^2\epsilon_0^2\int\delta n^2_y E_y^2\dif y\label{eq:deltahapprox}\\
    \frac{\delta\beta^2}{\beta^2} &\approx \frac{\int\delta n_z^2 E_z^2\dif y}
    {\int n_z^2 E_z^2\dif y}
    \xRightarrow{\delta n_z \text{ const. in AR}}
    \delta\beta\approx \frac{\beta\int_{\text{AR}}n_z E_z^2\dif y}
    {\int n_z^2 E_z^2\dif y} \delta n_z\equiv \Gamma\delta n_z
    \label{eq:confine}
\end{align}
\end{widetext}
where AR stands for the active region and as we will show, $\Gamma$ 
is the confinement factor when the non-perturbed material is Hermitian. 

This result is the optical version of the quantum mechanical treatment of 
Hamiltonian operators in \cite{PhysRevC.6.114}.

For QCLs the change in the refractive index within the active region derives
from the electrical dipoles between subbands, which is anisotropic
($\delta n_y^2 = \chi$ the electrical susceptibility from the dipole moment 
and $\delta n_z = 0$), so the approximation in Eq.~(\ref{eq:deltahapprox}) 
becomes exact. For a generic gain medium the perturbation difference is 
not necessarily of this form, like in a diode laser, where the gain is often 
isotropic ($\delta n_y = \delta n_z \neq 0$), 
this approximation is justified from the fact that $E_z$ is usually much larger 
than $E_x$ in a TM mode.

When we neglect the difference in group and phase velocities of the material, 
the gain of the medium is proportional to the imaginary part of the refractive
index: 
\begin{equation}
    I = I_0 \left|\e^{\mi n \omega z /c}\right|^2 = I_0\e^{g z}
    \quad\Rightarrow\quad g = -2\imag\frac{\omega}{c} n
\end{equation}
where $I$ is the optical power flow and $n$ is the complex refractive index, 
including the active gain. Similarly for a guided mode, 
$\modi{g} = -2\imag\frac{\omega}{c}\beta$. 
With $n^2 = n_z^2 + \chi$ and $\beta+\delta\beta$ given above,
\modi{the relationship between the the active material gain (given in forms of $\chi$) 
and the waveguide effective gain $g_\text{eff}$ induced by $\chi$} is given by:
\begin{align}
    &g = -2\imag \frac{\omega}{c}\sqrt{n_z^2+\chi}
     \approx -\frac{\omega}{n_zc}\imag\chi \\
    &g_{\text{eff}} = -2\imag\frac{\omega}{c}\delta\beta
    = -\imag\frac{\omega\beta\int_{\text{AR}} \chi E_z^2\dif y} 
    {c\int n_z^2 E_z^2\dif y} \label{eq:exactgain}
\end{align} 

We employ a linear response form of $g_{\text{eff}} = g\Gamma$ to define a 
confinement factor $\Gamma$, but in general this is possible only when the 
following are approximately true:
(a) the gain medium is uniform on the wavelength scale and linear,
i.e. $\chi$ does not depend on the electrical field and is therefore constant 
for the active region;
(b) the linear response does not mix the real and imaginary part of the perturbed 
refractive index, in other word, $\Gamma$ in Eq.~(\ref{eq:confine}) is real.

In the simplest case where the material is low loss, the matrix in 
Eq.~(\ref{eq:single_tranfer}) means that the fields at different parts of 
the waveguide are always in phase,
therefore, the confinement factor can be written in the following form 
that is more reminscent to the frequently used formula
$g_\text{eff}= \Gamma g$:
\begin{equation}
    \Gamma \approx \frac{(\real\beta)\int_{\text{AR}}n_z |E_z|^2\dif y}
    {\int n_z^2|E_z|^2\dif y} \label{eq:confinementreal}
\end{equation}

However, generally, the linear response of the waveguide effective gain 
for a perturbed bulk gain in the active region is not necessarily real, 
meaning $\Gamma$ is complex or the real part of $\chi$
has an effect on the imaginary part of $\beta$ and vice versa. 
This becomes more relevant when the device is working on a frequency that's 
off-resonance to the intrinsic frequency of the gain medium, where the 
Lorentzian shape introduces an out-of-phase component of the dipole oscillation 
and therefore an electrical susceptibility $\chi$ with both non-zero real and 
imaginary part (versus, when working on-resonance, $\chi$ is purely imaginary).

Comparing the above results and two frequently used formulas for the confinement 
factor:
\begin{align}
    &\Gamma_{\text{\cite{coldren2012diode}}} 
    = \frac{\int_{\text{AR}}n|E|^2\dif y}{\int n|E|^2\dif y}
    \label{eq:confinement_ne} \\
    &\Gamma_{\text{\cite{modeling}}}
    = \frac{\int_{\text{AR}} |E|^2\dif y}{\int |E|^2\dif y}
    \label{eq:confinement_e}
\end{align}
The difference is shown in Fig.~\ref{fig:device-gain} 
for the waveguide structure from \cite{firstgaas} and \cite{firstplasmon}
with different imaginary part of susceptibility in the active region,
where we can see that the widely used confinement factor formulas have a few 
percent error compared to the revised version solution, while our equation shows 
one to two orders of magnitude smaller relative error, 
particularly the Eq.~(\ref{eq:exactgain}) is the exact linear
term of the gain in active medium. 
\modi{It is worth mentioning that, in our context, 
the plasmonic boundary for the structure in 
Ref.~\cite{firstplasmon} shares the same physical model as a double-metal 
waveguide for THz QCLs, 
except that the latter are more sensitive to the electrical and optical properties of 
the metal, which in tern depends on the metal deposition process and which we lack more 
information of; yet we are expecting our proposed formula to show similar 
improvement for double-metal waveguide.}

\begin{figure*}[!htp]
\begin{minipage}[t]{0.5\linewidth}
\centering
\includegraphics[width=\linewidth]{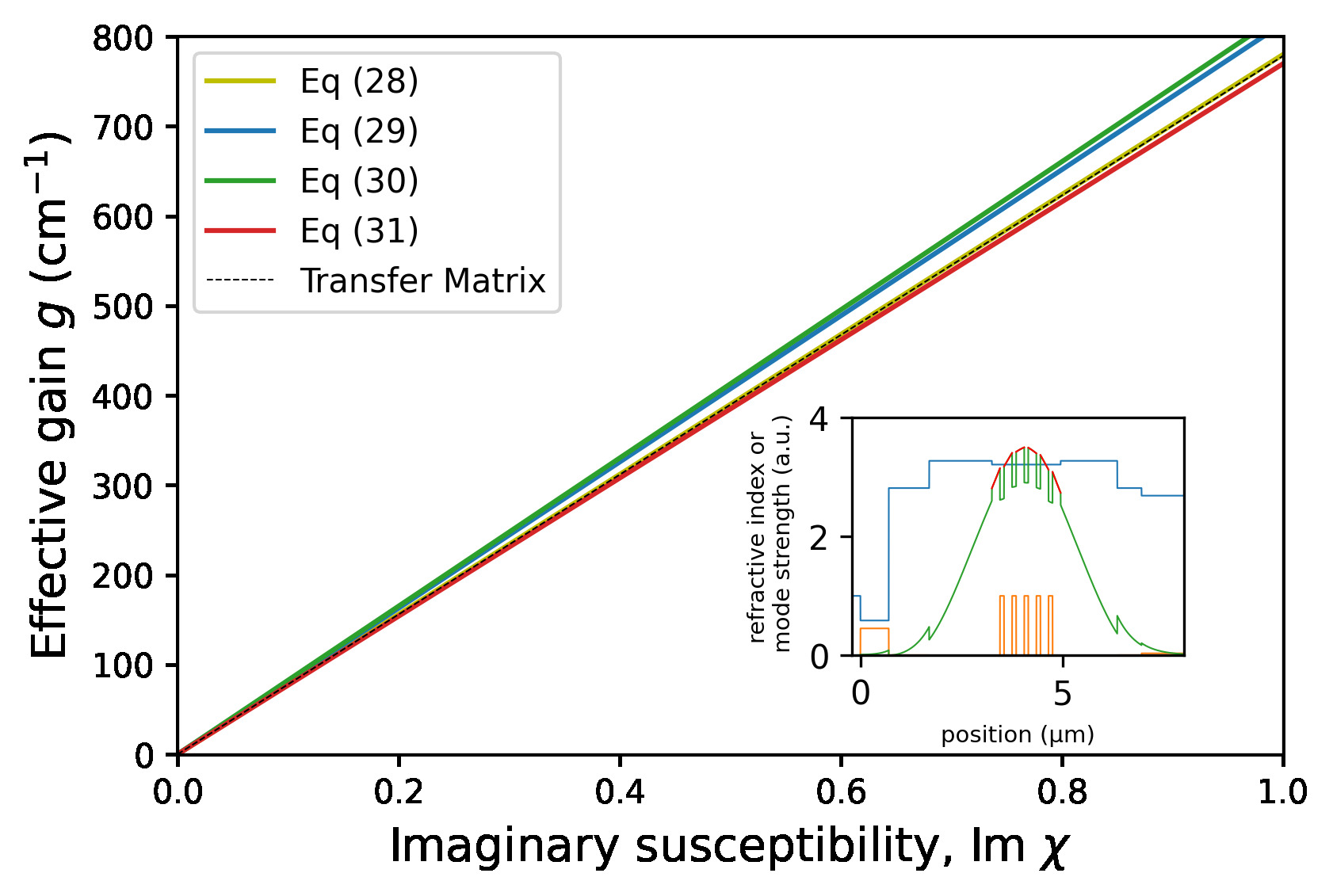}
\end{minipage}%
\begin{minipage}[t]{0.5\linewidth}
\centering
\includegraphics[width=\linewidth]{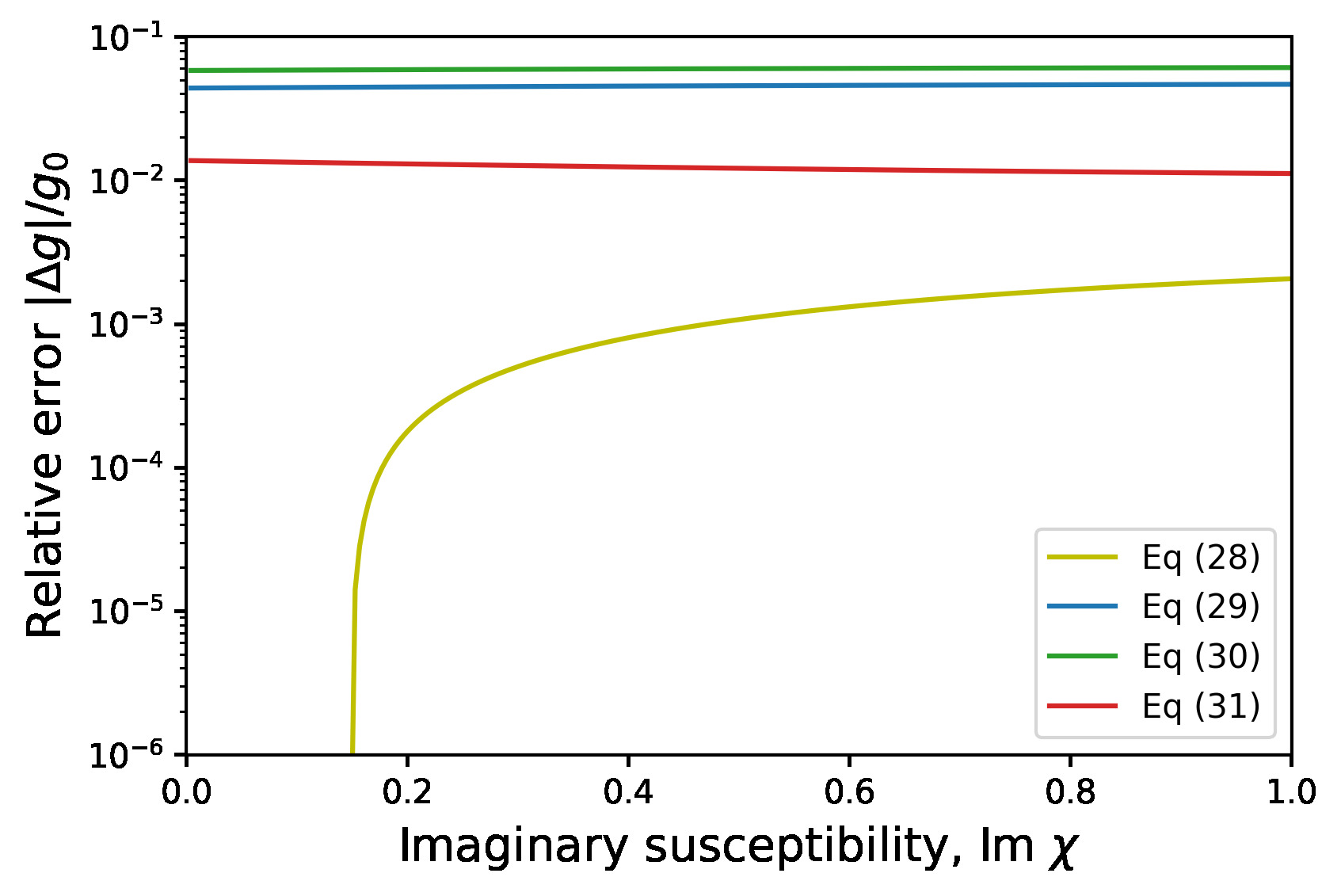}
\end{minipage}
\caption{The effective optical gain of the waveguide vs the imaginary 
susceptibility of the QC active material (\textbf{left}) and the relative 
error of the linear estimation (\textbf{right}), 
calculated from Eq.~(\ref{eq:exactgain}) and
the confinement factor defined in 
Eq.~(\ref{eq:confinementreal})--(\ref{eq:confinement_e}). 
The non-perturbed result (dashed black line) from the transfer matrix method is 
considered the exact result.
\text{Insets} are the waveguide structure for the calculation, 
where the blue lines and orange lines are the real and imaginary part of 
the refractive index respectively; 
green lines are the mode strength in arbitrary units, 
with the field in the active region colored red. 
\label{fig:alternating-gain}}
\end{figure*}

To show when the difference between Eq.~(\ref{eq:exactgain}) and 
Eq.~(\ref{eq:confinementreal}) is more significant, the comparison for a
structure with alternating QC gain and high-doped lossy material is shown in 
Fig.~\ref{fig:alternating-gain}. 
Such a structure may be of interest as a potential candidate for negative 
refractive index materials \cite{hoffman2007negative}. 

\section{Conclusion}
In summary, 
we have derived corrected formulas for the effective medium refractive index of 
the active region and the confinement factor 
for the purpose of QCL waveguide design.
The difference to commonly used formulas of the confinement factor and effective 
refractive index in prior literature is up to a few percent in a typical 
waveguide for QCLs, due to the inaccurate linear response, 
due to neglecting the anisotropic property or the non-Hermitian property 
of the QC materials. 
The difference may become large when there is highly lossy material inside
the device. 

The method in this work can in straight forward manner be extended to 
other optical devices.
By preserving the extra $E_z$ term in Eq.~(\ref{eq:deltahapprox}) and by 
modifying the field vector basis in the transfer matrix 
Eq.~(\ref{eq:single_tranfer}) as in \cite{Chilwell:84} the result can 
be easily generalized to any layered-structure active or passive optical devices 
for both TE and TM mode with isotropic or anisotropic gain.

\begin{acknowledgments}
We would like to thank Yezhezi Zhang for helpful discussion with the potential 
application of this work to active meta-materials.
This work was supported by the Intellectual Property Accelerator Fund received 
from Princeton University’s Office of the Dean for Research, 
and the Bede Liu Research Fund for Electrical Engineering received from 
Princeton University’s Department of Electrical and Computer Engineering.
\end{acknowledgments}

\section*{data availability statement}
Data sharing is not applicable to this article as no new data were created or analyzed in this
study.

\bibliography{optics}
\end{document}